\documentclass[fdp,a4paper,fleqn]{w-art}
\usepackage{times,cite,w-thm,amsfonts}
\theoremstyle{plain}

\theoremstyle{definition}

\usepackage[]{graphicx}
\newcommand{\eq}[1]{eq.\,(\ref{#1})}
\def\ii{\mathrm{i}}
\def\ee{\mathrm{e}}

\newcommand{\vev}[1]{\langle #1 \rangle}

\newcommand{\tr}{\mathrm{tr\,}}

\newcommand{\cE}{{\mathcal{E}}}

\newcommand{\cM}{{\mathcal{M}}}

\newcommand{\cO}{{\mathcal{O}}}

\newcommand{\cZ}{{\mathcal{Z}}}

\newcommand{\one}{{\rm 1\kern -.9mm l}}

\begin{document}


\title[Holographic couplings]{Non-perturbative gauge couplings from holography}

\author[M. Bill\`o]{Marco Bill\`o\inst{1}%
  \footnote{Corresponding author\quad E-mail:~\textsf{billo@to.infn.it},
            Phone: +39\,011\,670\,7213,
            Fax: +39\,011\,670\,7214.}}
\address[\inst{1}]{Dipartimento di Fisica, Universit\`a di Torino\\
and I.N.F.N. - sezione di Torino,
Via P. Giuria 1, I-10125 Torino, Italy}
\author[M. Frau]{Marialuisa Frau\inst{1}
}
\author[L. Giacone]{Luca Giacone\inst{1}
}
\author[A. Lerda]{Alberto Lerda\inst{2}
} 
\address[\inst{2}]{Dipartimento di Scienze e Innovazioni Tecnologiche, Universit\`a del Piemonte
Orientale\\
and I.N.F.N. - Gruppo Collegato di Alessandria, 
Viale T. Michel  11, I-15121 Alessandria, Italy}
\begin{abstract}
  We show how D-instanton corrections modify the dilaton-axion profile emitted
  by an O7/D7 system turning it into the non-singular F-theory background which corresponds
  to the effective coupling on a D3 probe. 
\end{abstract}

%% maketitle must follow the abstract.
\maketitle                   

\section{Introduction}

We consider the local limit of Type I$^\prime$ string theory with $N_f$ D7 branes close to
an O7 plane and study the profile of the corresponding dilaton-axion field $\tau$. 
At the perturbative level, 
$\tau$ is non-trivial and possesses logarithmic singularities at the 
orientifold and brane positions. These singularities are incompatible with its r\^ole as 
string coupling constant, and must therefore be resolved by non-perturbative effects, resulting 
into a non-singular F-theory background.
The $N_f=4$ case was considered long ago by A. Sen \cite{Sen:1996vd} who, based on the 
symmetries and monodromy properties of the Type I$^\prime$ configuration, suggested that the
exact dilaton-axion profile be given by the effective coupling
of the $4d$ $\mathcal{N}=2$ SYM theory with gauge group SU(2) and $N_f=4$ flavors,
as encoded in the corresponding Seiberg-Witten curve \cite{Seiberg:1994aj}.
This is the gauge theory supported by a probe D3 brane in the local Type I$^\prime$ model,
and thus the F-theory background can be interpreted as the gravity dual of the 
effective gauge coupling of the D3 brane world-volume theory \cite{Banks:1996nj}. 

Here we provide a microscopic description of how the \emph{exact} F-theory background arises 
when D-instantons are introduced in the D7/O7 system and show how they modify 
the source terms in the $\tau$ field equation \cite{Billo:2011uc}. 
The computation requires integrating over the D-instanton moduli space and
this is done via localization techniques \cite{Moore:1998et,Nekrasov:2002qd} that allow to obtain 
explicit results even when all instanton numbers contribute. In this way we demonstrate how the
non-perturbative corrections to the effective gauge coupling are incorporated in the dual
gravitational solution. The agreement of the exact dilaton-axion profile thus obtained with the
coupling constant of the D3 brane gauge theory
persists all the way down to $N_f=0$, which amounts to say that an O7 plane plus its D-instanton 
corrections represents the gravitational source for the gauge coupling of the pure 
$\mathrm{SU}(2)$ $\mathcal{N}=2$ SYM theory in $4d$.

The local configuration we consider contains $N_f$ D7 branes; 
the massless excitations of the D7/D7 open strings describe a gauge theory in eight dimensions and
the orientifold projection implies that its gauge group is $\mathrm{SO}(N_f)$.
These degrees of freedom can be assembled into an adjoint chiral superfield 
\begin{equation}
\label{Mdef}
 M=m+\theta \psi+ \frac12 \theta \gamma^{\mu\nu}\theta F_{\mu\nu}+\cdots~.
\end{equation}
A D3-brane (plus its orientifold image) in this background supports a 
four-dimensional $\mathrm{Sp}(1)\sim \mathrm{SU(2)}$ gauge theory with $N_f$ hypermultiplets, arising from the 
D3/D7 strings, and flavor group $\mathrm{SO}(N_f)$. 
For $N_f=4$ this theory has vanishing $\beta$-function, and we will mostly consider
this case, except in the last section.

The transverse space to the O7 plane and the D7 branes (parametrized by a complex coordinate $z$) 
corresponds to the Coulomb branch of the moduli space of the theory: placing the probe 
D3 brane in $z$ (and its image in $-z$) amounts to give a vacuum expectation value 
$\phi_{\mathrm{cl}} = (a,-a)$, with $a=\frac{z}{2\pi\alpha'}$, to the SU(2) complex adjoint scalar.
On the other hand, displacing the D7 branes in $z_i$ ($i=1,\ldots,N_f$) corresponds to
giving a vacuum expectation value 
\begin{equation}
m_{\mathrm{cl}} = (m_1,\ldots,m_{N_f},-m_1,\ldots,-m_{N_f})~,~~\mbox{~~with~~}\phantom{8}m_i = \frac{z_i}{2\pi\alpha'}~,
\label{vevm}
\end{equation}
to the $\mathrm{SO}(N_f)$ complex adjoint scalar $m$ of \eq{Mdef}.
In the D3 brane effective action, the $m_i$'s represent the masses of the hypermultiplets,
while the r\^ole of the complexified gauge coupling is played by the dilaton-axion field 
$\tau$
belonging to the closed string sector. Actually, $\tau$ is the first component of
a chiral scalar superfield $T$ in which all relevant massless closed strings degrees of freedom 
can be organized and which is schematically given by \cite{Schwarz:1983qr}
\begin{equation}
 \label{Tdef}
 T= \tau+\theta \lambda + \cdots + 2 \theta^8 \big(\partial^4 \bar{\tau} 
 + \cdots\big)~,
\end{equation}
where $\partial$ stands for $\frac{\partial}{\partial z}$.
Both the O7 plane and the D7 branes couple to $T$ and produce a non-trivial profile for it.
This fact allows to establish an explicit gauge/gravity relation: the effective 
coupling $\tau(a)$ of the SU(2) theory on the probe D3 brane is the dilaton-axion background
$\tau(z)$ produced by the D7/O7 system.
Such a background is na\"ively ({\it i.e.} perturbatively) 
singular but, as we will show in the next sections, it can be promoted to a full-fledged 
non-singular F-theory background by taking into account non-perturbative D-instanton corrections.
On the gauge theory side, this amounts to promote the perturbative SU(2) gauge coupling to the 
exact one encoded in the corresponding Seiberg-Witten curve. 

\section{The dilaton-axion profile}
\label{sec:axio-dilaton}
As we mentioned above, the D7 branes and the O7 plane act as sources for $\tau$, 
localized in the transverse directions. The classical perturbative dilaton-axion 
profile corresponding to $N_f=4$ D7 branes placed in $z_i$ is given by
\begin{equation}
\label{pertprof}
   2\pi\ii\,\tau_{\mathrm{cl}}(z)  = 2\pi\ii\,\tau_0 +\sum_{i=1}^{4} 
   \Bigl[\log \frac{z -z_i}{z} + \log \frac{z + z_i}{z}\Bigr]
 = 2\pi\ii\,\tau_0 
   - \sum_{\ell=1}^\infty
   \frac{(2\pi\alpha')^{2\ell}}{2\ell}\,\frac{\tr m_{\mathrm{cl}}^{2\ell}}{z^{2\ell}} ~,
\end{equation}
where in the second step we used \eq{vevm}.
This profile, which matches the 1-loop running of the gauge coupling of the
SU(2) SYM theory with $N_f=4$ flavors, can be obtained by computing the 1-point function of the
$\tau$ emission vertex with the boundary states of the D7 branes and the crosscap state of the O7 
plane. The dilaton-axion (\ref{pertprof}) solves the equation of motion 
$\square\tau_{\mathrm{cl}} = J_{\mathrm{cl}} \, \delta^2(z)$, where the classical current
\begin{equation}
 J_{\mathrm{cl}}= -2\ii\sum_{\ell=1}^\infty \frac{(2\pi\alpha')^{2\ell}}{(2\ell)!}\,
\tr m_{\mathrm{cl}}^{2\ell}
\,\,\partial^{2\ell}
\label{Jcl}
\end{equation}
arises from interactions on the D7 world-volume between $\tau$ and the SO(8) adjoint 
scalar $m$ when the latter is frozen to its vacuum expectation value (\ref{vevm}). 
Such interactions can be obtained from a source action of the form
\begin{equation}
\label{Ssource}
S_{\mathrm{source}} \,= \,-\frac{1}{(2\pi)^3\,(2\pi\alpha')^4}
\!\int d^{8}x \,J_{\mathrm{cl}}\, {\bar\tau}
\end{equation}
where the dimensionful coefficient is the ratio of the D7 brane tension and the 
gravitational coupling constant, which is the appropriate normalization for a D7 source action \cite{Billo:2011uc}. 
Using the superfields $M$ and $T$ of eq.s (\ref{Mdef}) and (\ref{Tdef}), we can easily realize 
that the above interactions can be derived from the following perturbative $8d$ prepotential
\begin{equation}
\label{intrprep}
     F_{\mathrm{cl}}(M,{T})  = 2\pi\ii\sum_{\ell}\frac{(2\pi\alpha')^{2\ell-4}}{(2\ell)!}
    \,\tr \,M^{2\ell}~\partial^{2\ell-4}\,T~.
\end{equation}
Comparing the corresponding classical action 
$S_{\mathrm{cl}} =\frac{1}{(2\pi)^4}\int d^8x\,d^8\theta~F_{\mathrm{cl}}(M,T)$
with the definition (\ref{Ssource}) of the source action, we thus obtain
\begin{equation}
\label{Jclis}
   J_{\mathrm{cl}}= - \frac{(2 \pi \alpha')^4}{2 \pi} 
   \frac{\delta F_{\mathrm{cl}}}{\delta(\theta^8 {\bar\tau})}\Big|_{T=\tau_0,M=m_{\mathrm{cl}}} 
   \equiv \, - \frac{(2 \pi \alpha')^4}{2 \pi} \,\bar\delta F_{\mathrm{cl}}
~,
\end{equation}
where we introduced the handy notation 
${\bar\delta}\, \star \,\equiv \, \frac{\delta\,\star}{\delta(\theta^8\bar\tau)}\big|_{{T=\tau_0},{M=m_{cl}}}$~.

{From} this analysis it is clear how one should proceed to obtain the complete dilaton-axion 
source $J$. 
One has first to promote the classical prepotential to the full one by including non-perturbative corrections, 
{\it i.e.} $F(M,T)= F_{\mathrm{cl}}(M,{T})+ F_{\mathrm{n.p.}}(M,T)$, and then
write, in analogy to \eq{Jclis}, $J =  - \frac{(2 \pi \alpha')^4}{2 \pi}\, \bar\delta F$.

The non-perturbative contribution to the prepotential arises when D-instantons are added 
to the D7/O7 system; in this case new types of excitations appear corresponding to open strings 
with at least one end-point on the instantonic branes, {\it{i.e.}} D(--1)/D(--1) or D(--1)/D7 strings. 
Due to the boundary conditions, these excitations do not describe dynamical
degrees of freedom but account instead for the instanton moduli, which we collectively denote
by $\cM_{(k)}$\,, where $k$ is the instanton number. Among them, one finds the coordinates
of the center of mass and their fermionic partners, which can be identified with the $8d$ superspace
coordinates $x$ and $\theta$, respectively. The interactions among the moduli are encoded in the
instanton action and can be computed systematically by string diagrams 
as described in Ref.s~\cite{Billo:2002hm,Green:2000ke}.
In the case at hand, the instanton action can be written as
\begin{equation}
\label{Sinst}
 S_{\mathrm{inst}}(\cM_{(k)}, M, {T}) = S(\cM_{(k)}) + S(\cM_{(k)},M) +   S(\cM_{(k)},T)
\end{equation}
where $S(\cM_{(k)})$ is the pure moduli action, which corresponds to the ADHM measure on the 
moduli space, $S(\cM_{(k)}, M)$ is the mixed moduli/gauge fields action and
finally $S(\cM_{(k)},T)$ is the mixed moduli/gravity action. 
Here we focus on the most relevant part for our goal, namely $S(\cM_{(k)},T)$, and refer to the
literature for the other terms \cite{Billo:2009di,Billo:2011uc}. 
To obtain $S(\cM_{(k)},T)$,
we compute mixed open/closed string disk diagrams involving instanton moduli and bulk fields. 
The simplest diagrams yield the ``classical'' instanton action $-2\pi\ii\, k\,{\tau}$ 
supersymmetrized by insertions of $\theta$ moduli, so that $\tau$ gets replaced by 
the superfield $T$, resulting in $-2\pi\ii\, k\,T$.
Other mixed diagrams contributing to $S(\cM_{(k)}, M)$ involve the bosonic modulus {$\chi$}, which 
is akin to the position of the D(--1)'s in the transverse space to the D7's, but
with anti-symmetric Chan-Paton indices due to the orientifold projection. Such diagrams turn out 
to be exactly computable, even if they involve an arbitrary (even) number of $\chi$ insertions.
Moreover, they are supersymmetrized by $\theta$-insertions that promote 
all $\tau$ occurrences to $T$.   
Altogether, the mixed {moduli}/{gravity} action is \cite{Billo:2011uc}
\begin{equation}
\label{Smix}
   S({\cM_{(k)}}, T)=
   -2\pi\ii\sum_{\ell=0}^{\infty}
   \frac{(2\pi\alpha')^{2\ell}}{(2\ell)!}\,
   {\tr(\chi^{2\ell})}\,\bar p^{2\ell}\,T~,
\end{equation} 
where $\bar p$ is the momentum conjugate to $z$. 

Given the complete instanton action (\ref{Sinst}), one can obtain the non-perturbative effective 
action on the D7 branes by performing an integral over the D(--1) moduli space, namely
\begin{equation}
\label{Snpint}
  S_{\mathrm{n.p.}}= \sum_k \int d\cM_{(k)} 
    \ee^{-S_{\mathrm{inst}}(\cM_{(k)}, {M}, {T})}
    = \int d^{8}x \,d^{8}\theta \, F_{\mathrm{n.p.}}({M},{T})~,
\end{equation}
where in the last step we have explicitly exhibited the integral over the superspace
coordinates $x$ and $\theta$ to define the non-perturbative prepotential. 
The latter therefore arises from
an integral over all remaining instanton moduli, also called centered moduli.
Such an integral can be explicitly computed using localization 
techniques \cite{Moore:1998et,Nekrasov:2002qd}. 
This amounts to select one of the preserved supercharges 
as a BRST charge $Q$ so that the instanton action (\ref{Sinst})
is $Q$-exact, and to organize the instanton moduli in BRST doublets so that the integral over 
them reduces to the evaluation of determinants around the fixed points of $Q$.
In order to have isolated fixed points, the instanton action 
must be deformed by suitable parameters (to be removed at the end) which in our set-up
arise from a particular RR graviphoton background \cite{Billo:2006jm,Billo:2009di}. 
Here, we will not delve into the details but simply recall the essential ingredients of the 
procedure.

One first introduces the $k$-instanton partition 
function $Z_k$ according to
\begin{equation}
 Z_k = \int d\cM_{(k)} \,\ee^{-S_{\mathrm{inst}}(\cM_{(k)}, M, {T};\cE)}~,
\label{Zk}
\end{equation}
where $\cE$ is the deformation parameter. Then, setting $q = \ee^{2\pi\ii\tau_0}$
and $Z_0=1$, one writes the gran-canonical partition function
$\cZ = \sum_{k=0}^\infty q^k Z_k$, from which one obtains the non-perturbative prepotential
\begin{equation}
\label{Fnpq}
    F_{\mathrm{n.p.}} = 
\lim_{\cE \to 0}\cE \log \mathcal Z = \sum_{k=1}^\infty q^k\,F_k~.
\end{equation}
For example, $F_1=\lim_{\cE \to 0}\cE Z_1$, 
$F_2=\lim_{\cE \to 0}\cE \big(Z_2-\frac{1}{2}Z_1^2\big)$ and so on.
In complete analogy with \eq{Jclis}, one then writes the instanton-induced source 
for the dilaton-axion as
${J_{\mathrm{n.p.}}} =  -\frac{(2\pi\alpha')^4}{2\pi}\,{\bar\delta} {F_{\mathrm{n.p}}}$, so that its
$q$-expansion involves the variations ${\bar\delta} F_k$, which in turn are related to 
the variations ${\bar\delta} Z_k$. For example, ${\bar\delta} F_1=\lim_{\cE \to 0}\cE {\bar\delta} Z_1$, 
${\bar\delta} F_2=\lim_{\cE \to 0}\cE \big({\bar\delta} Z_2-Z_1{\bar\delta} Z_1\big)$ and so on.
Given the explicit form (\ref{Smix}) of the moduli action, it readily follows that
\begin{equation}
\label{deltaZk}
    {\bar\delta}{Z_k} = 4\pi\ii \sum_{\ell=0}^\infty 
    (2\pi\alpha')^{2\ell}\,\bar p^{2\ell+4}\,{Z_k^{(2\ell)}}~,
\end{equation}
where we introduced the ``correlators'' of the $\chi$-moduli in the instanton 
matrix theory
\begin{equation}
\label{chicorr}
    {Z_k^{(2\ell)}} =\frac{1}{(2\ell)!}\,\int \!\!{d\cM_{(k)}} ~
    {\tr(\chi^{2\ell})}\,
    \ee^{-S_{\mathrm{inst}}(\cM_{(k)}, M, {T};\cE)}\,\Big|_{{T=\tau_0}\,,\,{M=m_{\mathrm{cl}}}}~.
\end{equation}
At the first two instanton numbers one finds
\begin{equation}
\begin{aligned}
 {\bar\delta} F_1&=4\pi\ii \sum_{\ell=0}^\infty 
    (2\pi\alpha')^{2\ell}\,\bar p^{2\ell+4}\lim_{\cE\to0}\cE Z_1^{(2\ell)}~,\\
 {\bar\delta} F_2&=4\pi\ii \sum_{\ell=0}^\infty 
    (2\pi\alpha')^{2\ell}\,\bar p^{2\ell+4}\lim_{\cE\to0}\cE \big(Z_2^{(2\ell)}-
Z_1Z_1^{(2\ell)}\big)~,
\end{aligned}
\label{f1f2}
\end{equation}
and similar expressions can be easily obtained for any $k$.
The same combinations of partition functions $Z_k$ and $\chi$-correlators ${Z_k^{(2\ell)}}$ 
appear in the computation of the D-instanton contributions to a rather different class 
of observables, namely the protected correlators $\vev{\tr m^J}$ forming the chiral ring 
of the SO(8) gauge theory defined in the $8d$ world-volume of the
D7 branes. The non-perturbative part of the chiral ring elements 
have a $q$-expansion, $\vev{\tr m^J}_{\mathrm{n.p.}} = \sum_{k=1}^\infty q^k\,\vev{\tr m^J}_k$, 
which can be explicitly computed using localization techniques as discussed in 
Ref.~\cite{Fucito:2009rs}. At the first two instanton numbers one finds
\begin{equation}
\lim_{\cE\to0}\cE Z_1^{(2\ell)} = \frac{(-1)^\ell}{(2\ell+4)!}\,\vev{\tr m^{(2\ell+4)}}_1
~,\quad
 \lim_{\cE\to0}\cE \big(Z_2^{(2\ell)}-
Z_1Z_1^{(2\ell)}\big) =\frac{(-1)^\ell}{(2\ell+4)!}\,\vev{\tr m^{(2\ell+4)}}_2
\label{m1m2}
\end{equation}
and so on. Using these results, we therefore find a very strict relation between the $\bar\delta$
variation of the prepotential and the non-perturbative SO(8) chiral 
ring \cite{Billo:2010mg,Fucito:2011kb,Billo':2011pr}, namely
\begin{equation}
\label{deltatochir}
    {\bar\delta} {F_k} =  4\pi\ii\sum_{\ell=0}^\infty 
    (2\pi\alpha')^{2\ell}\,\bar p^{2\ell+4}\frac{(-1)^\ell}{(2\ell+4)!}\,
    {\vev{\tr m^{2\ell+4}}}_{{k}}~,
\end{equation}
which, taking into account the fact that ${\vev{\tr m^{2}}}_{{k}}=0$ for all $k$, implies that  
\begin{equation}
\label{Jnp}
   {J_{\mathrm{n.p.}}} = -\frac{(2\pi\alpha')^4}{2\pi} \sum_{k=1}^\infty q^k\, 
   {\bar\delta}  {F_k} =
   -2\ii \sum_{\ell=1}^\infty (-1)^\ell\,
   \frac{(2\pi\alpha')^{2\ell}\,{\bar p}^{2\ell}}{(2\ell)!} \,\vev{\tr m^{2\ell}}_{\mathrm{n.p.}}
   ~.
\end{equation}
Adding this expression (rewritten in the $z$ coordinate space) to the classical term 
$J_{\mathrm{cl}}$ of \eq{Jclis} yields the complete source $J$.
Solving the field equation
$\square\,{\tau} = {J}\,\delta^2(z)$, we get then the exact dilaton-axion profile
\begin{equation}
    2\pi\ii\,{\tau(z)} 
    = 2\pi\ii\,{\tau_0}  - \sum_{\ell=1}^\infty
    \frac{(2\pi\alpha')^{2\ell}}{2\ell}\,
    \frac{{\vev{\tr m^{2\ell}}}}{z^{2\ell}} = 2\pi\ii\,{\tau_0} +  
{\Big\langle{\log\det \Big(1 - \frac{(2\pi\alpha')\,m}{{z}}\Big)}\Big\rangle}~.
\label{tauz}
\end{equation}
At the perturbative level, \eq{pertprof} expressed that fact that 
the quantities $\tr m_{{\mathrm{cl}}}^{2\ell}$ of the D7 theory act as a source for
the dilaton-axion. This source, however, is non-perturbatively corrected 
and the exact result is obtained by replacing the classical vacuum expectation values 
with the full quantum correlators in the D7-brane theory, namely
$\tr m_{{\mathrm{cl}}}^{2\ell} \,\equiv \,\tr \vev{m^{2\ell}} ~~\to~~\vev{\tr m^{2\ell}}$.
Furthermore, introducing the operator 
\begin{equation}
 \cO_\tau(z) = \tau_0+\frac{1}{2\pi\ii}\log\det 
\Big(1 - \frac{(2\pi\alpha')\,m}{{z}}\Big)
\label{Otau}
\end{equation}
we can rewrite \eq{tauz} as $\tau(z) = \big\langle \cO_\tau(z)\big\rangle$, which has the 
typical form of a holographic relation.

\section{Comparison with gauge theory results}
\label{sec:comparison}
As we said above, the chiral ring elements ${\vev{\tr m^{2\ell}}}$ 
are explicitly computable via {localization} and for the first few values of $\ell$ 
their instanton expansion can be found in Ref.~\cite{Fucito:2009rs}. Using these results in (\ref{tauz})
and parametrizing the transverse directions with $z = 2\pi\alpha'a$ in such a way that
all $\alpha'$ factors in the $\tau$ profile are reabsorbed, we get an expression 
$\tau(a)$ that, by direct comparison, can be seen to be exactly equal to
the large-$a$ expansion of the low-energy {effective coupling} of the SU(2) SYM theory 
with $N_f=4$ massive flavors, as derived from the Seiberg-Witten curve \cite{Billo:2010mg}.
We can therefore rephrase this result in the following relation
\begin{equation}
\label{tauza}
\tau_{\mathrm{sugra}}(z) \Leftrightarrow \tau_{\mathrm{gauge}}(a)
\end{equation}
where $a=\frac{z}{2\pi\alpha'}$ represents the Coulomb branch parameter.
It is interesting to remark that on the supergravity side 
the non-perturbative contributions to the dilaton-axion profile $\tau_{\mathrm{sugra}}(z)$ 
are due to ``exotic'' instanton configurations in the $8d$ world-volume theory on the D7 branes, 
while on the gauge theory side the non-perturbative effects in $\tau_{\mathrm{gauge}}(a)$ 
arise from standard gauge instantons in the $4d$ SYM theory. 
These two types of contributions agree because they actually have the same microscopic origin: 
in both cases they are due to D(--1) branes, which represent ``exotic'' instantons for the D7/O7 
system and ordinary instantons for the probe D3 brane supporting the $4d$ SYM theory.  

The relation (\ref{tauza}) can obviously be used in two ways. On the one hand, it can be used 
to read the gauge coupling constant of the
$4d$ SU(2) theory in terms of the quantum correlators in the $8d$ theory which gauges
its SO(8) flavor symmetry. This is the approach we have discussed so far. 
On the other hand, the relation (\ref{tauza}) can be used to read the $8d$ chiral ring elements 
in terms of the $4d$ gauge coupling. 
Actually this can be done in an exact way, {\it{i.e.}} to all orders in $q$.
In fact, using recursion relations of Matone type, it is possible 
to extract from the Seiberg-Witten curve the exact expression of any given coefficient of the 
expansion of $\tau_{\mathrm{gauge}}(a)$ in inverse powers of $a$.
For example, from the exact coefficient of $\frac{1}{a^4}$ we can deduce that
\begin{equation}
 \vev{\tr m^{4}} = E_2(q) \,R^2 - 6\theta_4^4(q)\,T_1+ 6\theta_2^4(q)\,T_2
\end{equation}
where $E_2$ is the Eisenstein series of (almost) weight 2, $\theta_2$ and
$\theta_4$ are Jacobi $\theta$-functions and $R$, $T_1$ and $T_2$ are the quadratic and quartic
SO(8) mass invariants (see Ref.~\cite{Billo':2011pr} for details). This expression resums the
instanton expansion $\sum_{k=1}^\infty  q^k \,\vev{\tr m^{4}}_k$ in which only the first few terms  
were known by direct evaluation, and can be generalized to all other chiral ring elements.

\section{The pure SU(2) theory}
\label{sec:pure}
The gauge/gravity relation (\ref{tauza}) can be established also when some or all flavors are
decoupled to recover the asymptotically free theories with $N_f=3,2,1,0$. In particular, 
from the gauge theory side one can reach the pure SU(2) case by sending $q\to 0$ and 
$m_i\to \infty$, while keeping the combination $q\,m_1m_2m_3m_4 \equiv \Lambda^4$ finite. 
$\Lambda^4$ is the dimensionful counting parameter in the instanton expansion which replaces the 
dimensionless $q$ of the superconformal $N_f=4$ theory; in other words $\Lambda$ can be interpreted 
as the dynamically generated scale of the pure SU(2) theory.
{From} the supergravity side the decoupling of the flavors corresponds to sending all D7 branes far away
from the origin, or equivalently to evaluate the dilaton-axion $\tau(z)$ at a
$z$ much smaller than the D7 brane positions in such a way that only the orientifold O7 plane acts 
as a source for $\tau$. In this case we can therefore repeat the
same steps described in Section \ref{sec:axio-dilaton} and evaluate the dilaton-axion field 
emitted by just the O7 plane. 

At the non-perturbative level when $k$ D-instantons are put on the orientifold plane, the main 
difference with respect to the case with the D7's is that the spectrum of instanton moduli 
contains only neutral moduli corresponding to open strings of type D(--1)/D(--1). 
The rest of the derivation remains as before. In particular eq.s (\ref{Zk}) and (\ref{chicorr}) 
are well-defined even in the absence of the D7's, and thus formulas like (\ref{f1f2}) 
continue to hold. For example, one finds 
\begin{equation}
 \lim_{\cE\to0}\cE Z_1^{(2\ell)}=-\frac{12}{4!}\,\delta_{\ell,0}~,\quad\quad
\lim_{\cE\to0}\cE \big(Z_2^{(2\ell)}-
Z_1Z_1^{(2\ell)}\big)=-\frac{105}{4\cdot 8!}\,\delta_{\ell,2}
\label{z1z2}
\end{equation}
which are the analogue of \eq{m1m2} for the pure SU(2) theory. Of course the interpretation 
as chiral ring elements in a flavor theory is no longer possible since there are no flavors. 
Using these results and their generalizations at higher instanton numbers $k$, we can 
still find the variations $\bar\delta F_k$ of the prepotential and obtain from these the non-perturbative 
source current $J_{\mathrm{n.p.}}$, whose first instanton terms are
\begin{equation}
 J_{\mathrm{n.p.}}=2\ii\Big[\Lambda^4\,(2\pi\alpha' \bar p)^4\,\frac{12}{4!}+
\Lambda^8\,(2\pi\alpha' \bar p)^8\,\frac{105}{4\cdot 8!}+\cdots\,\Big]~.
\label{jnp0}
\end{equation}
Solving the field equation 
$\square\tau_{\mathrm{n.p.}}(z)= J_{\mathrm{n.p.}}\delta^2(z)$ and expressing the result in terms of 
$a=\frac{z}{2\pi\alpha'}$, we get 
\begin{equation}
 2\pi\ii\,\tau_{\mathrm{n.p.}}(a)=3\,\frac{\Lambda^4}{a^4}+
\frac{105}{32}\,\frac{\Lambda^8}{a^8}+\cdots
\end{equation}
which exactly coincides with the first two instanton contributions as derived from the Seiberg-Witten
curve \cite{Finnell:1995dr,Nekrasov:2002qd,Flume:2002az}. After adding the 
perturbative piece $2\pi\ii\,\tau_0-4\log\big(\frac{4a^2}{\Lambda^2}\big)$, we can therefore 
conclude that the full dilaton-axion profile sourced by an O7 plane plus its D-instantons 
completely agrees with the exact coupling of the pure $\mathrm{SU}(2)$ $\mathcal{N}=2$ SYM 
theory in $d=4$.

\vspace{8pt}
\noindent
This work was supported in part by the MIUR-PRIN contract 2009-KHZKRX.

\providecommand{\WileyBibTextsc}{}
\let\textsc\WileyBibTextsc
\providecommand{\othercit}{}
\providecommand{\jr}[1]{#1}
\providecommand{\etal}{~et~al.}

\end{document}